\documentclass[sigconf]{acmart}

\copyrightyear{2021} \acmYear{2021}
\setcopyright{acmlicensed}\acmConference[ARES 2021]{The 16th
	InternationalConference on Availability, Reliability and Security}{August
	17--20,2021}{Vienna, Austria}\acmBooktitle{The 16th International 
	Conference on
	Availability, Reliability and Security (ARES 2021), August 17--20, 2021, 
	Vienna,
	Austria}\acmPrice{15.00}\acmDOI{10.1145/3465481.3465769}\acmISBN{978-1-4503-9051-4/21/08}

\usepackage{amsmath,amsfonts}
\usepackage{newtxmath}
\usepackage{booktabs}
\usepackage{tabularx}
\usepackage{enumitem}
\usepackage{hyperref}
\usepackage{acronym}
\usepackage{caption}
\usepackage{subcaption}

\usepackage{xspace}
\newcommand*{\eg}{e.g.\@\xspace}
\newcommand*{\ie}{i.e.\@\xspace}

\newacro{RRT}{Randomized Response Technique}
\newacro{Bf}{Bloom filter}
\newacro{DP}{Differential Privacy}
\newacro{LDP}{Local Differential Privacy}

\usepackage{xcolor}
\definecolor{sron0}{HTML}{332288}
\definecolor{sron1}{HTML}{88CCEE}
\definecolor{sron2}{HTML}{117733}
\definecolor{sron3}{HTML}{DDCC77}
\definecolor{sron4}{HTML}{CC6677}
\definecolor{sron5}{HTML}{AA4499}

\begin{document}

\title{Self-Determined Reciprocal Recommender System\\with Strong Privacy
Guarantees\vspace{1ex}}

 \thanks{The authors acknowledge the financial support by the Federal Ministry
 of Education and Research of Germany in the framework of AnyPPA
 (project no. 16KIS0909).}

\author{Saskia Nu\~nez von Voigt}
\email{saskia.nunezvonvoigt@tu-berlin.de}
\affiliation{\institution{Distributed Security Infrastructures\\Technische Universit\"at
  Berlin}
  \city{Berlin}
  \country{Germany}
  \vspace{2ex}
}

\author{Erik Daniel}
\email{erik.daniel@tu-berlin.de}
\affiliation{\institution{Distributed Security Infrastructures\\Technische Universit\"at
  Berlin}
  \city{Berlin}
  \country{Germany}
  \vspace{2ex}
}

\author{Florian Tschorsch}
\email{florian.tschorsch@tu-berlin.de}
\affiliation{\institution{Distributed Security Infrastructures\\Technische Universit\"at
  Berlin}
  \city{Berlin}
  \country{Germany}
  \vspace{2ex}
}

\begin{abstract}
Recommender systems are widely used.
Usually, recommender systems are based on a centralized client-server
architecture.
However, this approach implies drawbacks regarding the privacy of users.
In this paper, we propose a distributed reciprocal recommender system with strong,
self-determined privacy guarantees, i.e., local differential privacy.
More precisely, users randomize their profiles locally and exchange them via a
peer-to-peer network. Recommendations are then computed and ranked locally by
estimating similarities between profiles.
We evaluate recommendation accuracy of a job recommender system and demonstrate
that our method provides acceptable utility under strong privacy requirements.
\end{abstract}

\maketitle
\pagestyle{plain}

\section{Introduction}
Recommender systems support users in their search for relevant information,
products, and services by presenting items to a user that match her
preferences or needs.
In a reciprocal recommender system, users are recommended to each other.
These recommendations are different from one-sided product recommendations,
where products are recommended to the user but not vice versa.
A typical example for reciprocal recommender systems are dating platforms,
where people search for a matching partner~\cite{XiaLSC15}.
Other areas that use two-sided recommender systems are mentoring systems
and job recommendations---the use case of our
work.

Most existing reciprocal recommender systems
provide a platform that connects users to each other.
The systems are based on a centralized client-server architecture.
In order to receive recommendations,
users disclose a significant amount of personal data
to create a user profile, which is collected and stored on a server.
This approach poses a serious threat to users' privacy and data control.

In this paper,
we propose a content-based distributed recommendation approach.
We develop a distributed recommender system
in which users exchange their profiles directly
to calculate similarities and to receive recommendations.
By following this approach, users determine whom to share the data with
and therefore gain control over their data.
To this end, we use Bloom filters~\cite{bloom1970space} to encode user profiles
and apply the randomized response technique~(RRT)
to provide strong privacy guarantees,
\ie, differential privacy~\cite{Dwork2006}.
In order to exchange user profiles, we avoid a central architecture.
At the same time, we believe that a synchronous peer-to-peer (P2P) network
would not be practically viable.
Instead, we propose to exchange data via a P2P data network,
specifically the InterPlanetary File System (IPFS)~\cite{benet2014ipfs},
which offers asynchronous data access.

We apply and discuss our approach in the context of
a P2P job marketplace, which we use as running example.
Not only since the EU's General Data Protection Regulation~(GDPR),
processing personal data is regulated.
In particular, personal data are considered
sensitive and require special protection.
Hence, we believe that a job marketplace is an ideal use case
as it has two sides (recruiters and candidates)
and emphasizes the data protection needs.

In our evaluation, we expose the impact of the privacy guarantees on
utility.
In particular, we show that comparable utility can be achieved
using a Bloom filter to encode profiles, instead of a binary vector.
Even for strong privacy guarantees ($\varepsilon<10$),
parameters can be found
which recommend more than 75\,\% of the top 20 candidates accurately
with high precision for the top ranked candidates.

The main contribution of our work is to provide
a \emph{distributed reciprocal recommender system
that is self-determined and offers strong privacy guarantees.}
The remainder of this paper is organized as follows:
We describe our running example, the P2P job marketplace in
Section~\ref{sec:usecase}.
In Section~\ref{sec:recommend}, we introduce our privacy-preserving 
recommendation
system and provide information about the privacy guarantees.
Subsequently, we describe data distribution in Section~\ref{sec:network}.
In Section~\ref{sec:eval}, we evaluate our system regarding the privacy-utility 
trade off.
We review related work in Section~\ref{sec:relwork}
and conclude the paper in Section~\ref{sec:conclusion}.

\section{Use case: Job Marketplace}
\label{sec:usecase}
Recommender systems are usually implemented as centralized platforms.
This however makes the platform less self-determined,
prone to data breaches, and implies privacy risks.
In order to emphasize this situation,
we use a job marketplace as running example.
According to the GDPR,
processing Human Resource~(HR) data is allowed under restricted circumstances
and increased means of data protection only.
In particular, it is difficult for consent to be freely given by job candidates,
meaning that it is unlikely to provide a valid basis for processing HR data.
Therefore, the marketplace is a relevant example
for a distributed system with self-determined data control.

\paragraph{Example: Two-sided Job Marketplace}
In recent years, recruiting and selection processes
are part of the success of the organization's future growth and
retention of employees.
The marketplace is divided into two parts: job recommendation and
candidate recommendation~\cite{MalinowskiKWW06}.
Matching jobs with candidates is based on similarities of the corresponding
profiles. These profiles may include features that consist of keywords
extracted from job requirements and candidate skills/preferences.
That is, job candidates receive suitable job recommendations
and recruiters receive candidate recommendations for a job.
More specifically, top ranked jobs or the top ranked candidates that best fit
the candidate profile or job profile, respectively, are returned.
The marketplace can therefore be classified as
a reciprocal recommender system.

On the marketplace, recruiters provide a job description
including a job profile with requirements and
necessary meta-data for the job application.
Candidates, interested in using the marketplace,
can provide their profile including skill set or interests,
or only receive job profiles to calculate possible recommendations.
Since job profiles contain no personal information,
we assume that they can be made publicly available.
Moreover, we assume that marketplace users (candidates and recruiters) are
globally distributed.
The system is most suitable for
global inter- or intra-company recruitment.

Our approach concentrates on the recommendation process.
The recruitment procedure takes place outside of the marketplace, thus
it is out of scope of this work.
 \section{Recommendation Approach}
\label{sec:recommend}
In this section, we describe the profile representation and the similarity score
for our privacy-preserving content-based recommendation approach.
Content-based recommendations use features of items and compute similarities
between items to recommend similar items that users have preferred in the past.
That means for the proposed marketplace,
we assume that candidates prefer jobs that are similar to their previous
jobs.

\subsection{Profile Representation}
The job profiles as well as the candidate profiles are based on keywords
extracted from job requirements and candidate skills, respectively.
This set of keywords can be represented as a vector.

While in principle the overall set of keywords is finite and well defined,
the number of possible keywords can be very large and new keywords may not
appear in new jobs. Therefore, we propose to use a probabilistic data
structure to represent profiles, keep the memory footprint small, and to
be able to handle unknown keywords.

Hence, we represent the set of keywords
in a Bloom filter~\cite{bloom1970space}.
Bloom filters have been introduced to approximate membership queries
but can also serve as item representation for recommender systems~\cite{PozoCMM16}.
The data structure of a Bloom filter consists of a bit array
of a fixed length~$m$ using~$k$ different hash functions~\cite{bloom1970space}.
At the beginning, the Bloom filter is empty,
\ie, all bits in the array are set to~$0$.
The hash function~$h_i(x)$ with~$i \in \{1,\dots,k\}$ takes an input~$x$
and determines a position in the bit array that is accordingly set to~$1$.
Membership of a value~$x$ can be tested by repeating the process
and checking all respective positions $h_i(x)$ in the array.
If there is at least one bit set to $0$,
the value~$x$ is not member of the Bloom filter.
If all corresponding bits are set to $1$,
the value~$x$ may be a member of the Bloom filter, but we cannot be certain.
Due to bit collisions, influenced by the parameters~$m$ and~$k$,
false positives can occur.

\begin{figure}
	\centering
	\includegraphics{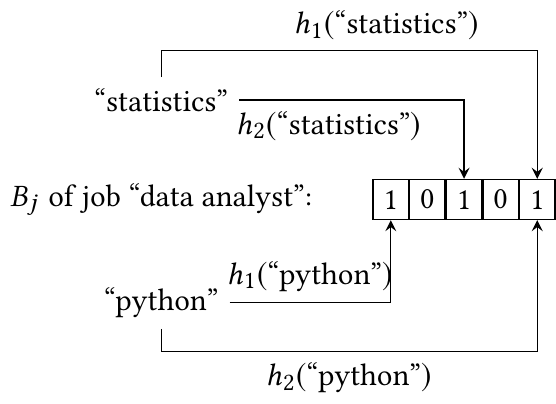}
	\vspace{-1ex}
	\caption{Process of adding keywords to a Bloom filter with $k=2$ and
		$m=5$.}
	\label{fig:BloomFilter}
	\vspace{-2ex}
\end{figure}

For our use case, each job/candidate profile consists of a Bloom filter
of the same length and with the same number of hash functions.
In Figure~\ref{fig:BloomFilter}, we sketch the procedure of adding the keywords
to the Bloom filter.
Assume for example the job profile~"{data analyst}"
has the keywords~"{statistics}" and "{python}".
These keywords are hashed and the corresponding bits are set to $1$.

\subsection{Job Recommendation}
In order to obtain recommendations,
candidates can derive personalized recommendations locally
without disclosing their profile.
To this end, we use the cosine similarity as metric
as it is directly applicable to vectors
and widely used in content-based recommender systems~\cite{PuglisiPFR15}.
That is, candidates download the Bloom filter of a job offering~$B_j$
and compare it to their own Bloom filter~$B_c$.
They calculate the cosine similarity of the two Bloom filters, given by
\begin{equation}
	\cos\theta = \frac{B_j\cdot B_c}{\Vert B_j \Vert \Vert B_c\Vert},
	\label{eq:cos}
\end{equation}
where the numerator is the scalar product.

Since our proposed recommender system is self-determined,
candidates can identify the top $N$ jobs with the highest similarity.
Additionally, candidates can also set a threshold value, \ie,
jobs with a similarity higher than a customizable value are recommended.

\subsection{Candidate Recommendation}
A recruiter needs candidate profiles
to obtain the top $N$ candidates for her job offering.
We assume that the recruiter has received candidate profiles
in the representation of a Bloom filter
as described in the previous section.

Due to the false positive rate of a Bloom filter,
there is a certain uncertainty that provides some privacy.
For example in Figure~\ref{fig:BloomFilter},
a hash value of "{statistics}" and "{python}"
are both mapped to the same bit (5th bit).
While this provides some privacy,
it is unable to guarantee strong privacy~\cite{BianchiBL12}.
For example, an adversary could still determine
that a profile does \emph{not} have certain keywords
by inspecting the bits set to $0$.

Our goal is therefore to offer strong privacy guarantees for profiles,
\ie, differential privacy which enables plausible deniability~\cite{Dwork2006}.
Differential privacy is a strong privacy definition,
providing privacy guarantees to the input of a computation
regardless of the amount of background knowledge of an adversary.
Local differential privacy~(LDP) satisfies differential privacy
in the local setting.
In other words, a data collector is not trusted
and therefore candidates perturb their profile locally.

A function~$f$ provides $\epsilon$-differential privacy~\cite{Dwork2006}
if for all neighboring pairs of profiles $B$ and $B'$
and all $S\subseteq \text{Range}(f)$ satisfy
\begin{equation}
	P[f(B) \in S] \leq \mathrm{e}^{\epsilon} P[f(B') \in S]\text{.}
	\label{eq:dp}
\end{equation}
In other words, the result of~$f$ should be similar independent of the
input, \ie, it is irrelevant whether a candidate reports $B$ or $B'$.

LDP is guaranteed by flipping the bits of the Bloom filter based on the RRT,
a method used for surveys~\cite{warner1965randomized}.
With a probability~$p$, the bit of the Bloom filter is flipped,
\ie, a~$1$ changes to~$0$ and vice versa a~$0$ to~$1$.
Otherwise, the bit is not changed and remains the same.
In other words, each bit in the Bloom filter ($0$ \emph{and} $1$) yields
plausible deniability as it remains unclear whether a set bit is a result of
randomization or truly corresponds to a certain keyword.
Differential privacy is guaranteed with
$p=\frac{1}{1+e^{\epsilon/k}}$.

A candidate distributes the differentially-private Bloom filter $B_c'$
to recruiters.
The recruiter can then calculate the scalar product
between the perturbed candidate's profile $B_c'$ and the job profile $B_j$
to determine the similarity between the candidate and job profiles.
Due to the randomization, however,
he obtains a perturbed scalar product~$\tilde{SP}$.
To obtain an unbiased cosine similarity, the scalar product is
corrected
with the number of~$1$
in the candidates perturbed Bloom filter $\Vert B_c'\Vert_1$
according to~\cite{alaggan2012blip} by
\begin{equation}
	\hat{SP} = \frac{\tilde{SP}-p\Vert B_c'\Vert_1}{1-2p},
\end{equation}
The cosine similarity is then calculated
with the corrected~$\hat{SP}$ and Equation~(\ref{eq:cos}).
For more details, we refer the reader to \cite{alaggan2012blip}.
Note that correcting the scalar product only removes noise from the similarity,
not from the candidate Bloom filters,
effectively preserving privacy.

\begin{figure}[t]
	\centering
	\includegraphics{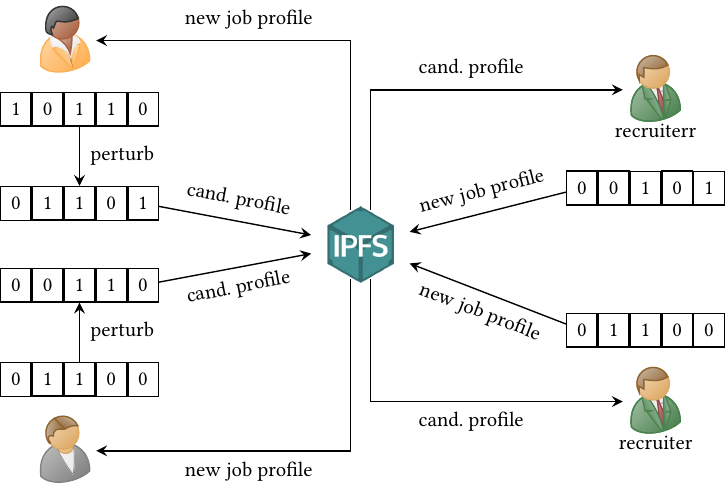}
	\caption{Privacy-preserving recommender system where profiles are
		exchanged via IPFS.}
	\label{fig:system}
	\vspace{-2ex}
\end{figure} \section{Content Distribution \& Storage}\label{sec:network}
The recommendation is calculated locally.
Therefore, candidates and recruiters need to exchange
candidate and job profiles, respectively.
Recruiters and candidates effectively build a P2P network.
In a naive approach, the P2P network needs to be synchronous
to enable direct exchange between recruiters and candidates.
While we might be able to assume that recruiters are available at all times,
candidates can go offline unexpectedly.

We therefore propose to use the InterPlanetary File System~(IPFS)
as distributed storage layer,
which we illustrate in Figure~\ref{fig:system}.
IPFS is a data network~\cite{daniel2021ipfs},
offering distributed data storage and sharing.
IPFS allows to distribute candidate and job profiles
asynchronously without introducing a central server.
IPFS therefore combines the best of both worlds
and users gain more control over their data.
Data is stored only on a user's device
and only replicated on demand.

IPFS uses the \emph{libp2p} library for its network layer and
the \emph{Bitswap} protocol~\cite{de2021accelerating} for exchanging data.
Files are split into blocks.
The blocks are used to build a Merkle Directed Acyclic Graph~(DAG).
Blocks are content-addressed and exchanged based on their Content Identifier~(CID).
The CID is derived from the content. Exchanging a file requires the root CID of the Merkle DAG.
Blocks contain information of their children
and acquiring the DAG results in the file.
Blocks are found by requesting neighbors or querying
a Kademlia-based Distributed Hash Table.

For our use case, we face three main challenges:
ensuring profile availability,
job and candidate profile discovery,
and access control.

\subsection{Profile Availability}
In order to ensure the data is available even after a node leaves the network,
data has to be replicated by multiple peers of the network.
In IPFS, data is replicated passively, by volunteers or cache-based.
Initially, data is made available by a single node only (the data source),
and will be available as long as the node remains online.
In addition, cache-based replication improves availability of popular content.
In order to ensure asynchronous availability, however, data needs to be actively replicated.
For active replication, there are two possibilities:
Filecoin~\cite{protocollabs2017filecoin} and IPFS Cluster.

Filecoin is a blockchain-based incentive layer,
where storage and retrieval of data is compensated via the filecoin token~(FIL).
In a job marketplace, this generally might be a feasible option:
recruiters pay a fee to make the job profiles available,
increasing the range of possible candidates.
Candidates pay the fee to make their profiles available
to increase their consideration for job offerings.
At the time of writing storing data in Filecoin
costs around $4.8\,pFIL/GiB$ for one epoch ($30\,s$)\footnote{https://file.app/ 
(2021-06-03)}
and the minimum duration for storage deals is $180\,d$.
The costs are therefore low with roughly
$2.5\,\mu FIL$ to store $1\,GiB$ ($1\,FIL\approx 72\,\$$)\footnote{https://coinmarketcap.com/en/currencies/filecoin/ (2021-06-03)}
and negligible for our use case.
Nevertheless, we believe that using Filecoin is more suitable for
an inter-company marketplace than an intra-company marketplace.
In an intra-company marketplace, it might be questionable,
who would be willing to pay for storage.
In general though, the main disadvantage is that using Filecoin introduces a blockchain
and can therefore introduce additional privacy problems.
A blockchain is an immutable public ledger,
and therefore reveals information about the pseudonymous participants.

Instead, IPFS Cluster provides the possibility for pinset orchestration.
The cluster ensures the availability of files
by managing a certain degree of replication (the pinset) between the cluster peers.
It actively replicates data, even when nodes leave the cluster.
The cluster peers build an overlay network,
separate from the IPFS overlay network,
and the replication is independent from the IPFS network.
Hence, recruiters build a cluster to increase the availability,
ensuring candidates can acquire available job profiles.
Additionally, the cluster can replicate candidate profiles,
if candidates give their consent.
Since recruiters have a strong interest to make job profiles constantly available,
IPFS Cluster is a viable solution to offer asynchronous,
yet distributed and self-determined, data sharing.

\subsection{Profile Discovery}
In order to query candidate and job profiles,
marketplace users need to know the respective CIDs of profiles. Specifically, data is addressed by the root CIDs of the Merkle DAG.
Since CIDs are based on the content, changing data changes the CID
and CIDs are in general not human-readable.
As a result, we need a mechanism to inform marketplace users
of existing job and candidate profiles.

To this end, we utilize the Publish/Subscribe~(pubsub) architecture
from \emph{libp2p} to announce CIDs.
In the network, peers subscribe to a topic,
\ie, job or candidate profiles,
and are informed of published content.
That is, recruiters publish their job profile in IPFS
and distribute the respective root CID in the network,
where subscribed candidates receive new CIDs
and use it to retrieve the profile.
Similarly, candidates can publish their profile and
recruiters receive the candidate profiles.
The information is disseminated in the network via gossiping.

While pubsub ensures that active users receive new messages,
new users also want to receive information about relevant profiles.
Therefore, a bootstrapping mechanism for CIDs is necessary.
To this end, we maintain a list of CIDs.
After creating a new profile, the list is updated and new data can be identified.
Due to content-addressing of CIDs,
the list's CID changes with every update.
In order to address this issue, we suggest to use IPFS's name service,
\ie, Interplanetary Name System~(IPNS) that maps the hash of a public key to a CID.
That is, changing content changes the mapping,
effectively informing about the availability of new job profiles.
The maintenance of the file itself will be managed by the IPFS Cluster.
If candidates do not want to share their data
with potentially everyone on the marketplace,
they can share their data deliberately using an out-of-band channel,
\eg, sending an e-mail with the CID.

\subsection{Access Control}
A downside of IPFS is the lack of built-in data access control.
Once shared, data subjects lack control over replication and access of data.
Moreover, there is no dedicated way to delete data.
In general, everyone can request all blocks and
curious users can observe requested/announced CIDs.
Since we consider job offerings public anyway,
the lack of access control is not a problem.
However, candidate profiles contain personal data such as name
and therefore need additional protection.
We suggest to use encryption of candidate profiles to maintain confidentiality.
The decryption key is then used as a way to gain access
by sharing it only deliberately via other channels with selected individuals.
Some research proposes a blockchain for managing access
and sharing decryption keys~\cite{wang2018blockchain}.
 \section{Evaluation}\label{sec:eval}
We perform our evaluation on a real dataset\footnote{\url{https://www.kaggle.com/c/job-recommendation/data}}
provided by the online employment website Careerbuilder.
The dataset contains, inter alia, information about job postings, candidates,
and their application history.
For our training set we randomly select 10,000 job postings from the
job dataset.
The candidate profiles are generated from the application history.
To this end, we selected all users with at least five applied jobs
that are in the job posting dataset as well.
Our test dataset for recommending candidates consist of new jobs of the training
set that none of the candidates have in their application history.
In total, we have 10,000 jobs, 128 candidate profiles,
and 7,175 open jobs profiles.

We assume that the requirements for a job offering are vectorized.
Therefore, we extract keywords from job titles, descriptions, and requirements
by using TF-IDF (term frequency-inverse document frequency), a commonly used scheme.
Keywords that occur frequently in one job~(term frequency), but rarely
in other job profiles~(inverse document frequency) have a higher score
and are more likely to be relevant to describe a job.
For each job, we used all keywords. In total, each job is on average described
by 176~keywords~(min 2~keywords; max 588~keywords).

In the experiment, we evaluate the candidate recommendation.
More specifically, for each job offering,
we compute the cosine similarity for each applied job in the candidates profile.
In order to generate candidate recommendation,
we rank the candidates according to their mean similarity of their applied jobs.

In our evaluation, we aim at comparing
the set of computed recommendations of the binary vector model
with a Bloom filter model (bf) and a Bloom filter model
with differential privacy guarantees~(bf-DP).
The binary vector uses a binary representation of the keywords of each job
and represents our ground truth.
In the other two models, keywords are added to Bloom filters.
Additionally, in bf-DP,
the Bloom filter is perturbed according to an~$\varepsilon$.

We evaluate the utility of our recommendations by considering a scenario where
the recommendation list contains the top $N$ candidates.
Therefore, we adopt precision@N as utility metric since it is widely used to
evaluate recommender systems \cite{PuglisiPFR15,XiaLSC15}.
Precision@N is the number of relevant candidates within the top~$N$
divided by~$N$.
Typically, for utility, precision@N is evaluated along with recall@N,
which is the number of relevant candidates within the top~$N$
divided by the number of relevant candidates.
Since we consider the top~$N$ candidates recommended by the binary vector model 
as relevant, the number of relevant candidates is also $N$,
hence precision@N and recall@N are identical.
That is, we measure the utility only in terms of precision@N.

A drawback of precision is that the order of candidates is not
taken into account.
Therefore, we additionally use the average precision that is defined as the average
of precisions at each rank for a relevant candidate up to rank~$N$.
For example, consider a set of 5 recommended candidates, where the candidates
at position 1,4 and 5 are relevant because they are also in the set recommended
by the binary model. The average precision of this list is then given by
$(1+0.5+0.6)/3=0.7$. An average precision of $1.0$ indicates that all relevant
candidates are at the first positions.

\subsection{Parameter Selection}
To implement our recommender system, we need to specify
a number of parameters. For the Bloom filters, length
$m$ and the number of hash functions $k$ must be specified.
To this end, we measure the mean precision@20 of 50 randomly selected jobs over
10 runs with $\varepsilon=\ln3$ to show how $k$ and $m$
affect the utility of bf-DP.

\begin{figure}[tb]
	\centering
	\includegraphics{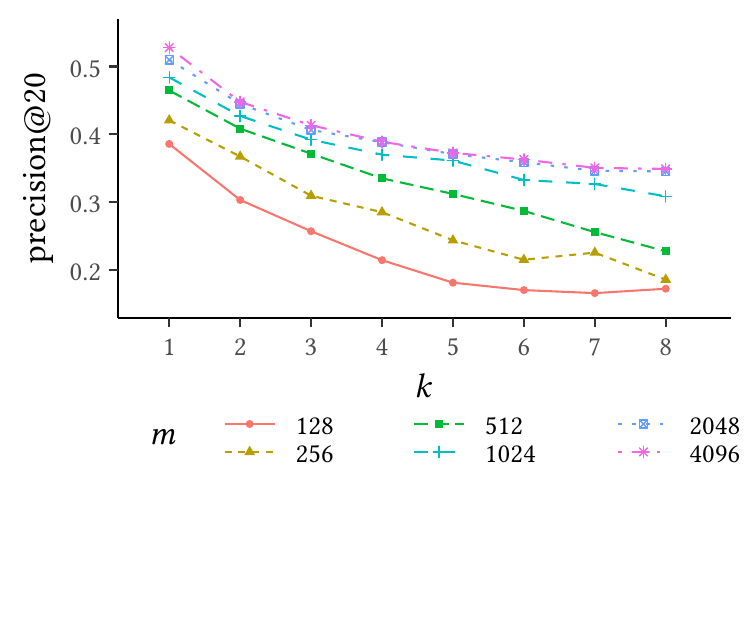}
	\vspace{-5em}
	\caption{Mean precision of bf-DP with $\varepsilon=\ln3$ considering the
		top 20 candidates (50 jobs over 10 runs) in dependence of Bloom
		filter length~$m$ and number of hash functions~$k$.}
	\label{fig:trendparam}
	\vspace{-2ex}
\end{figure}

In Figure~\ref{fig:trendparam}, we plotted the mean precision@20 with $k$ on the
x-axis and varying values for $m$.
The results generally indicate that a longer Bloom filter
leads to more utility and with more hash functions $k$ the precision decreases.
We obtain the highest precision with $k=1$ and $m=4096$.
Therefore, we set the Bloom filter length $m=4096$,
which corresponds to the length of the binary vector.

\subsection{Privacy-Utility Trade off}
\begin{figure*}[tb]
	\centering
	\begin{subfigure}[b]{0.45\textwidth}
		\centering
		\includegraphics{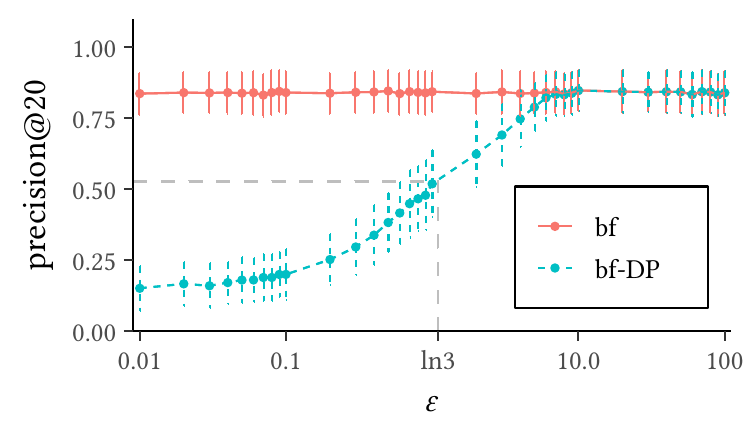}
		\vspace{-2ex}
		\caption{Mean precision@20 as utility metric.}
		\label{fig:utilityprec}
	\end{subfigure}
	\hfill
	\begin{subfigure}[b]{0.45\textwidth}
		\centering
		\includegraphics{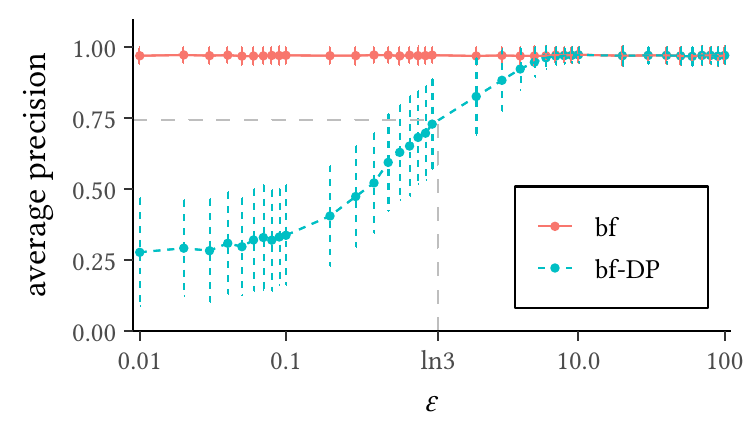}
		\vspace{-2ex}
		\caption{Mean average precision as utility metric. }
		\label{fig:utilitymap}
	\end{subfigure}
	\caption{Utility evaluation of bf and bf-DP with $m=4096$ and $k=1$ for 50
		jobs (10 runs) varying $\varepsilon$.}
	\label{fig:utility}
	\vspace{-2ex}
\end{figure*}

We quantify the cost of privacy by comparing bf
with bf-DP.
Please note that bf,
does not satisfy the definition of differential privacy.
In Figure~\ref{fig:utility}, we present the results for precision@20
(Figure~\ref{fig:utilityprec}) and average precision 
(Figure~\ref{fig:utilitymap})
versus the privacy loss parameter~$\varepsilon$ for the two models.
For all results, we show the mean of 50 randomly selected jobs over 10 runs;
standard deviation is used to draw the error bars.
Since only bf-DP depends on~$\varepsilon$,
the utility changes with varying~$\varepsilon$.
As expected, the precision@20 as well as the average precision of bf-DP
increases with higher~$\varepsilon$.
For a typical $\varepsilon=\ln3$~\cite{DworkR14},
the precision of bf-DP is below bf.
However, for~$\varepsilon>7$, the precision@20 of bf-dp approaches
the precision of bf.
Even for a small $\varepsilon=\ln3$, the average precision is $0.75$, \ie,
a high proportion of relevant candidates
appear early in the recommendation list.

\subsection{Discussion}
Our results demonstrate that Bloom filter can be used to generate
recommendations with strong privacy guarantees. With an $\varepsilon=4$, we
recommend approximately 75\,\% of the top 20 candidates correctly with an
average precision of 0.93.
By ensuring LDP, we provide a reciprocal recommender
system that is self-determined.
Candidates can decide whether to provide their profile.
For the sake of clarity, in our evaluation we assumed the same privacy
guarantee ($\varepsilon$) for each candidate.
However, since the unbiased similarity is computed
independently for each user, it is possible that each candidate chooses their
own desired privacy guarantee. This makes our system more self-determining.

When collecting data over time, the time series can leak information.
If noise is repeatedly added to the same or only slightly modified profiles,
the noise component can be averaged over a period of time and thus eliminated.
This becomes relevant when a candidates profile changes
or if a candidate has several similar applied jobs that are independently
perturbed.
Therefore, repeatable collection of profiles
requires an adjustment of the RRT.
Otherwise, a profile can be disclosed.
Future work should consider longitudinal attacks
and make appropriate adjustments such as memoization~\cite{ErlingssonPK14}.

Compared to bf-DP, bf provides almost the same accuracy as the binary vector.
Since a Bloom filter already provides certain deniability due to bit collisions,
we recommend using a Bloom filter instead of a binary vector.

Compared to a centralized approach, the distributed approach
has some disadvantages in performance and storage.
Due to splitting files into blocks by IPFS
and the construction of a Merkle DAG, a storage overhead is introduced.
Furthermore, profiles are replicated on multiple devices increasing overall storage requirements.
The usage of Bloom filters reduces the size of stored and transferred data,
making the data small enough to fit into one block.
Since the data fits into one block the additional transfer time
due to traversing and finding the DAG can be omitted.
An advantage of P2P systems is the self-scalability of data.
However, the small data size reduces the benefits due to replication.
Future work will concentrate on analyzing and minimizing possible storage and
transfer overhead.
 \section{Related Work}\label{sec:relwork}

Various approaches have been proposed for privacy-preserving recommender
systems.
They are based on cryptographic primitives~\cite{ErkinVTL12,MelisDC16}
or data
perturbation~\cite{FriedmanBK16,LiuLZLLZZ17,McSherryM09,PolatD03,SainiRJ19,ShinKSX18}.
The recommendations, however, are generated using collaborative filtering,
which uses preferences of many users to make recommendations.
While collaborative filtering is often preferred in terms of accuracy,
we use content-based filtering as it provides user-specific classification.
That is, recommendations are based on past user preferences,
which suits our distributed self-determined approach better.
Since it additionally considers the content,
it is particularly relevant for reciprocal recommender systems
and the application domain considered in this
work~\cite{LopsJMBK19,MalinowskiKWW06}.

Privacy-preserving content-based recommender systems are mostly concerned
with targeted advertising~\cite{Wang201821},
where privacy is achieved through anonymization or pseudonymization.
However, in a reciprocal recommender system, the use of these approaches would
not be feasible since the recommendation is two-sided and therefore the
candidates must also be known.
Puglisi et al.~\cite{PuglisiPFR15} propose a privacy-preserving content-based
recommender
by injecting arbitrary keywords into the user profile.
In contrast to Puglisi et al.,
we guarantee LDP for profiles.

Usually, differentially private recommender systems
are designed using a centralized
architecture~\cite{LiuLZLLZZ17,MelisDC16,SainiRJ19}.
We instead propose a distributed approach
that offers self-determined privacy guarantees,
using IPFS.
In the literature, IPFS is often used in combination with blockchains~\cite{wang2018blockchain}
or to store and exchange data~\cite{muralidharan2019interplanetary,ali2017iot}.
We use it for content distribution, enabling asynchronous data access.
 \section{Conclusion}\label{sec:conclusion}
In this paper, we present a distributed reciprocal recommendation system.
We use Bloom filters to allow local computation of recommendations with
differential privacy guarantees.
We show that Bloom filter as profile representation provide
reasonable utility under strong privacy guarantees.
Utilizing IPFS, we allow asynchronous exchange of data increasing users'
control over their data.

\bibliographystyle{ACM-Reference-Format}
\bibliography{local}


\begin{thebibliography}{28}


\ifx \showCODEN    \undefined \def \showCODEN     #1{\unskip}     \fi
\ifx \showDOI      \undefined \def \showDOI       #1{#1}\fi
\ifx \showISBNx    \undefined \def \showISBNx     #1{\unskip}     \fi
\ifx \showISBNxiii \undefined \def \showISBNxiii  #1{\unskip}     \fi
\ifx \showISSN     \undefined \def \showISSN      #1{\unskip}     \fi
\ifx \showLCCN     \undefined \def \showLCCN      #1{\unskip}     \fi
\ifx \shownote     \undefined \def \shownote      #1{#1}          \fi
\ifx \showarticletitle \undefined \def \showarticletitle #1{#1}   \fi
\ifx \showURL      \undefined \def \showURL       {\relax}        \fi
\providecommand\bibfield[2]{#2}
\providecommand\bibinfo[2]{#2}
\providecommand\natexlab[1]{#1}
\providecommand\showeprint[2][]{arXiv:#2}

\bibitem[\protect\citeauthoryear{Alaggan, Gambs, and Kermarrec}{Alaggan
  et~al\mbox{.}}{2012}]%
        {alaggan2012blip}
\bibfield{author}{\bibinfo{person}{Mohammad Alaggan},
  \bibinfo{person}{S\'{e}bastien Gambs}, {and} \bibinfo{person}{Anne-Marie
  Kermarrec}.} \bibinfo{year}{2012}\natexlab{}.
\newblock \showarticletitle{{BLIP: non-interactive differentially-private
  similarity computation on bloom filters}}. In
  \bibinfo{booktitle}{\emph{SSS'~12: Proceedings of the 14th International
  Symposium Stabilization, Safety, and Security of Distributed Systems}}.
  \bibinfo{publisher}{Springer}, \bibinfo{pages}{202--216}.
\newblock


\bibitem[\protect\citeauthoryear{Ali, Dolui, and Antonelli}{Ali
  et~al\mbox{.}}{2017}]%
        {ali2017iot}
\bibfield{author}{\bibinfo{person}{Muhammad~Salek Ali},
  \bibinfo{person}{Koustabh Dolui}, {and} \bibinfo{person}{Fabio Antonelli}.}
  \bibinfo{year}{2017}\natexlab{}.
\newblock \showarticletitle{IoT Data Privacy via Blockchains and {IPFS}}. In
  \bibinfo{booktitle}{\emph{IOT~'17: Proceedings of the 7th International
  Conference on the Internet of Things}}. \bibinfo{pages}{14:1--14:7}.
\newblock


\bibitem[\protect\citeauthoryear{Benet}{Benet}{2014}]%
        {benet2014ipfs}
\bibfield{author}{\bibinfo{person}{Juan Benet}.}
  \bibinfo{year}{2014}\natexlab{}.
\newblock \bibinfo{booktitle}{\emph{{IPFS} - Content Addressed, Versioned,
  {P2P} File System (DRAFT 3)}}.
\newblock \bibinfo{type}{{T}echnical {R}eport}. \bibinfo{institution}{Protocol
  Labs}.
\newblock


\bibitem[\protect\citeauthoryear{Bianchi, Bracciale, and Loreti}{Bianchi
  et~al\mbox{.}}{2012}]%
        {BianchiBL12}
\bibfield{author}{\bibinfo{person}{Giuseppe Bianchi}, \bibinfo{person}{Lorenzo
  Bracciale}, {and} \bibinfo{person}{Pierpaolo Loreti}.}
  \bibinfo{year}{2012}\natexlab{}.
\newblock \showarticletitle{"Better Than Nothing" Privacy with Bloom Filters:
  To What Extent?}. In \bibinfo{booktitle}{\emph{PSD~'12: Proceedings of the
  International Conference on Privacy in Statistical Databases}},
  Vol.~\bibinfo{volume}{7556}. \bibinfo{publisher}{Springer},
  \bibinfo{pages}{348--363}.
\newblock


\bibitem[\protect\citeauthoryear{Bloom}{Bloom}{1970}]%
        {bloom1970space}
\bibfield{author}{\bibinfo{person}{Burton~H Bloom}.}
  \bibinfo{year}{1970}\natexlab{}.
\newblock \showarticletitle{{Space/time trade-offs in hash coding with
  allowable errors}}.
\newblock \bibinfo{journal}{\emph{Commun. ACM}} \bibinfo{volume}{13},
  \bibinfo{number}{7} (\bibinfo{year}{1970}), \bibinfo{pages}{422--426}.
\newblock


\bibitem[\protect\citeauthoryear{Daniel and Tschorsch}{Daniel and
  Tschorsch}{2021}]%
        {daniel2021ipfs}
\bibfield{author}{\bibinfo{person}{Erik Daniel} {and} \bibinfo{person}{Florian
  Tschorsch}.} \bibinfo{year}{2021}\natexlab{}.
\newblock \showarticletitle{IPFS and Friends: A Qualitative Comparison of Next
  Generation Peer-to-Peer Data Networks}.
\newblock \bibinfo{journal}{\emph{arXiv preprint arXiv:2102.12737}}
  (\bibinfo{year}{2021}).
\newblock


\bibitem[\protect\citeauthoryear{De~la Rocha, Dias, and Psaras}{De~la Rocha
  et~al\mbox{.}}{2021}]%
        {de2021accelerating}
\bibfield{author}{\bibinfo{person}{Alfonso De~la Rocha}, \bibinfo{person}{David
  Dias}, {and} \bibinfo{person}{Yiannis Psaras}.}
  \bibinfo{year}{2021}\natexlab{}.
\newblock \showarticletitle{Accelerating Content Routing with Bitswap: A
  Multi-Path File Transfer Protocol in IPFS and Filecoin}.
\newblock  (\bibinfo{year}{2021}), \bibinfo{pages}{11}.
\newblock


\bibitem[\protect\citeauthoryear{Dwork, McSherry, Nissim, and Smith}{Dwork
  et~al\mbox{.}}{2006}]%
        {Dwork2006}
\bibfield{author}{\bibinfo{person}{Cynthia Dwork}, \bibinfo{person}{Frank
  McSherry}, \bibinfo{person}{Kobbi Nissim}, {and} \bibinfo{person}{Adam
  Smith}.} \bibinfo{year}{2006}\natexlab{}.
\newblock \showarticletitle{{Calibrating noise to sensitivity in private data
  analysis}}. In \bibinfo{booktitle}{\emph{TCC~'06: Proceedings of the 3rd
  Theory of Cryptography Conference}}. Springer, \bibinfo{pages}{265--284}.
\newblock


\bibitem[\protect\citeauthoryear{Dwork and Roth}{Dwork and Roth}{2014}]%
        {DworkR14}
\bibfield{author}{\bibinfo{person}{Cynthia Dwork} {and} \bibinfo{person}{Aaron
  Roth}.} \bibinfo{year}{2014}\natexlab{}.
\newblock \showarticletitle{The Algorithmic Foundations of Differential
  Privacy}.
\newblock \bibinfo{journal}{\emph{Foundations and Trends in Theoretical
  Computer Science}} \bibinfo{volume}{9}, \bibinfo{number}{3-4}
  (\bibinfo{year}{2014}), \bibinfo{pages}{211--407}.
\newblock


\bibitem[\protect\citeauthoryear{Erkin, Veugen, Toft, and Lagendijk}{Erkin
  et~al\mbox{.}}{2012}]%
        {ErkinVTL12}
\bibfield{author}{\bibinfo{person}{Zekeriya Erkin}, \bibinfo{person}{Thijs
  Veugen}, \bibinfo{person}{Tomas Toft}, {and} \bibinfo{person}{Reginald~L.
  Lagendijk}.} \bibinfo{year}{2012}\natexlab{}.
\newblock \showarticletitle{Generating Private Recommendations Efficiently
  Using Homomorphic Encryption and Data Packing}.
\newblock \bibinfo{journal}{\emph{{IEEE} Trans. Inf. Forensics Secur.}}
  \bibinfo{volume}{7}, \bibinfo{number}{3} (\bibinfo{year}{2012}),
  \bibinfo{pages}{1053--1066}.
\newblock


\bibitem[\protect\citeauthoryear{Erlingsson, Pihur, and Korolova}{Erlingsson
  et~al\mbox{.}}{2014}]%
        {ErlingssonPK14}
\bibfield{author}{\bibinfo{person}{\'{U}lfar Erlingsson},
  \bibinfo{person}{Vasyl Pihur}, {and} \bibinfo{person}{Aleksandra Korolova}.}
  \bibinfo{year}{2014}\natexlab{}.
\newblock \showarticletitle{{RAPPOR:} Randomized Aggregatable
  Privacy-Preserving Ordinal Response}. In \bibinfo{booktitle}{\emph{CCS~'14:
  Proceedings of the 2014 {ACM} {SIGSAC} Conference on Computer and
  Communications Security}}. \bibinfo{publisher}{ACM},
  \bibinfo{pages}{1054--1067}.
\newblock


\bibitem[\protect\citeauthoryear{Friedman, Berkovsky, and
  K{\^{a}}afar}{Friedman et~al\mbox{.}}{2016}]%
        {FriedmanBK16}
\bibfield{author}{\bibinfo{person}{Arik Friedman}, \bibinfo{person}{Shlomo
  Berkovsky}, {and} \bibinfo{person}{Mohamed~Ali K{\^{a}}afar}.}
  \bibinfo{year}{2016}\natexlab{}.
\newblock \showarticletitle{A differential privacy framework for matrix
  factorization recommender systems}.
\newblock \bibinfo{journal}{\emph{User Model. User Adapt. Interact.}}
  \bibinfo{volume}{26}, \bibinfo{number}{5} (\bibinfo{year}{2016}),
  \bibinfo{pages}{425--458}.
\newblock


\bibitem[\protect\citeauthoryear{Liu, Liu, Zhang, Li, Liu, Zhao, and Zhou}{Liu
  et~al\mbox{.}}{2017}]%
        {LiuLZLLZZ17}
\bibfield{author}{\bibinfo{person}{Xiao Liu}, \bibinfo{person}{An Liu},
  \bibinfo{person}{Xiangliang Zhang}, \bibinfo{person}{Zhixu Li},
  \bibinfo{person}{Guanfeng Liu}, \bibinfo{person}{Lei Zhao}, {and}
  \bibinfo{person}{Xiaofang Zhou}.} \bibinfo{year}{2017}\natexlab{}.
\newblock \showarticletitle{When Differential Privacy Meets Randomized
  Perturbation: {A} Hybrid Approach for Privacy-Preserving Recommender System}.
  In \bibinfo{booktitle}{\emph{{DASFAA}~'17: Proceedings of the 22nd
  International Conference on Database Systems for Advanced Applications}},
  Vol.~\bibinfo{volume}{10177}. \bibinfo{publisher}{Springer},
  \bibinfo{pages}{576--591}.
\newblock


\bibitem[\protect\citeauthoryear{Lops, Jannach, Musto, Bogers, and Koolen}{Lops
  et~al\mbox{.}}{2019}]%
        {LopsJMBK19}
\bibfield{author}{\bibinfo{person}{Pasquale Lops}, \bibinfo{person}{Dietmar
  Jannach}, \bibinfo{person}{Cataldo Musto}, \bibinfo{person}{Toine Bogers},
  {and} \bibinfo{person}{Marijn Koolen}.} \bibinfo{year}{2019}\natexlab{}.
\newblock \showarticletitle{Trends in content-based recommendation - Preface to
  the special issue on Recommender systems based on rich item descriptions}.
\newblock \bibinfo{journal}{\emph{User Model. User Adapt. Interact.}}
  \bibinfo{volume}{29}, \bibinfo{number}{2} (\bibinfo{year}{2019}),
  \bibinfo{pages}{239--249}.
\newblock


\bibitem[\protect\citeauthoryear{Malinowski, Keim, Wendt, and
  Weitzel}{Malinowski et~al\mbox{.}}{2006}]%
        {MalinowskiKWW06}
\bibfield{author}{\bibinfo{person}{Jochen Malinowski}, \bibinfo{person}{Tobias
  Keim}, \bibinfo{person}{Oliver Wendt}, {and} \bibinfo{person}{Tim Weitzel}.}
  \bibinfo{year}{2006}\natexlab{}.
\newblock \showarticletitle{Matching People and Jobs: A Bilateral
  Recommendation Approach}. In \bibinfo{booktitle}{\emph{HICSS-39~'06:
  Proceedings of the 39th Hawaii International International Conference on
  Systems Science}}. \bibinfo{publisher}{{IEEE} Computer Society}.
\newblock


\bibitem[\protect\citeauthoryear{McSherry and Mironov}{McSherry and
  Mironov}{2009}]%
        {McSherryM09}
\bibfield{author}{\bibinfo{person}{Frank McSherry} {and} \bibinfo{person}{Ilya
  Mironov}.} \bibinfo{year}{2009}\natexlab{}.
\newblock \showarticletitle{Differentially Private Recommender Systems:
  Building Privacy into the Netflix Prize Contenders}. In
  \bibinfo{booktitle}{\emph{SIGKDD~'09: Proceedings of the 15th {ACM}
  International Conference on Knowledge Discovery and Data Mining}}.
  \bibinfo{publisher}{{ACM}}, \bibinfo{pages}{627--636}.
\newblock


\bibitem[\protect\citeauthoryear{Melis, Danezis, and Cristofaro}{Melis
  et~al\mbox{.}}{2016}]%
        {MelisDC16}
\bibfield{author}{\bibinfo{person}{Luca Melis}, \bibinfo{person}{George
  Danezis}, {and} \bibinfo{person}{Emiliano~De Cristofaro}.}
  \bibinfo{year}{2016}\natexlab{}.
\newblock \showarticletitle{Efficient Private Statistics with Succinct
  Sketches}. In \bibinfo{booktitle}{\emph{NDSS~'16: Proceedings of the 23rd
  Annual Network and Distributed System Security Symposium}}.
  \bibinfo{publisher}{The Internet Society}.
\newblock


\bibitem[\protect\citeauthoryear{Muralidharan and Ko}{Muralidharan and
  Ko}{2019}]%
        {muralidharan2019interplanetary}
\bibfield{author}{\bibinfo{person}{Shapna Muralidharan} {and}
  \bibinfo{person}{Heedong Ko}.} \bibinfo{year}{2019}\natexlab{}.
\newblock \showarticletitle{An InterPlanetary File System {(IPFS)} based IoT
  framework}. In \bibinfo{booktitle}{\emph{ICCE~'19: Proceedings of the 37th
  {IEEE} International Conference on Consumer Electronics}}.
  \bibinfo{pages}{1--2}.
\newblock


\bibitem[\protect\citeauthoryear{Polat and Du}{Polat and Du}{2003}]%
        {PolatD03}
\bibfield{author}{\bibinfo{person}{Huseyin Polat} {and}
  \bibinfo{person}{Wenliang Du}.} \bibinfo{year}{2003}\natexlab{}.
\newblock \showarticletitle{Privacy-Preserving Collaborative Filtering Using
  Randomized Perturbation Techniques}. In \bibinfo{booktitle}{\emph{ICDM~'03:
  Proceedings of the 3rd {IEEE} International Conference on Data Mining}}.
  \bibinfo{publisher}{{IEEE} Computer Society}, \bibinfo{pages}{625--628}.
\newblock


\bibitem[\protect\citeauthoryear{Pozo, Chiky, Meziane, and M\'{e}tais}{Pozo
  et~al\mbox{.}}{2016}]%
        {PozoCMM16}
\bibfield{author}{\bibinfo{person}{Manuel Pozo}, \bibinfo{person}{Raja Chiky},
  \bibinfo{person}{Farid Meziane}, {and} \bibinfo{person}{Elisabeth
  M\'{e}tais}.} \bibinfo{year}{2016}\natexlab{}.
\newblock \showarticletitle{An item/user representation for recommender systems
  based on bloom filters}. In \bibinfo{booktitle}{\emph{RCIS~'16: Proceedings
  of the 10th {IEEE} International Conference on Research Challenges in
  Information Science}}. \bibinfo{publisher}{IEEE}, \bibinfo{pages}{1--12}.
\newblock


\bibitem[\protect\citeauthoryear{{Protocol Labs}}{{Protocol Labs}}{2017}]%
        {protocollabs2017filecoin}
\bibfield{author}{\bibinfo{person}{{Protocol Labs}}.}
  \bibinfo{year}{2017}\natexlab{}.
\newblock \bibinfo{booktitle}{\emph{Filecoin: A Decentralized Storage
  Network}}.
\newblock \bibinfo{type}{{T}echnical {R}eport}. \bibinfo{institution}{Protocol
  Labs}.
\newblock


\bibitem[\protect\citeauthoryear{Puglisi, Parra-Arnau, Forn{\'{e}}, and
  Rebollo-Monedero}{Puglisi et~al\mbox{.}}{2015}]%
        {PuglisiPFR15}
\bibfield{author}{\bibinfo{person}{Silvia Puglisi}, \bibinfo{person}{Javier
  Parra-Arnau}, \bibinfo{person}{Jordi Forn{\'{e}}}, {and}
  \bibinfo{person}{David Rebollo-Monedero}.} \bibinfo{year}{2015}\natexlab{}.
\newblock \showarticletitle{On content-based recommendation and user privacy in
  social-tagging systems}.
\newblock \bibinfo{journal}{\emph{Comput. Stand. Interfaces}}
  \bibinfo{volume}{41} (\bibinfo{year}{2015}), \bibinfo{pages}{17--27}.
\newblock


\bibitem[\protect\citeauthoryear{Saini, Rusu, and Johnston}{Saini
  et~al\mbox{.}}{2019}]%
        {SainiRJ19}
\bibfield{author}{\bibinfo{person}{Amar Saini}, \bibinfo{person}{Florin Rusu},
  {and} \bibinfo{person}{Andrew Johnston}.} \bibinfo{year}{2019}\natexlab{}.
\newblock \showarticletitle{PrivateJobMatch: a privacy-oriented deferred
  multi-match recommender system for stable employment}. In
  \bibinfo{booktitle}{\emph{RecSys~'19: Proceedings of the 13th {ACM}
  Conference on Recommender Systems}}. \bibinfo{publisher}{{ACM}},
  \bibinfo{pages}{87--95}.
\newblock


\bibitem[\protect\citeauthoryear{Shin, Kim, Shin, and Xiao}{Shin
  et~al\mbox{.}}{2018}]%
        {ShinKSX18}
\bibfield{author}{\bibinfo{person}{Hyejin Shin}, \bibinfo{person}{Sungwook
  Kim}, \bibinfo{person}{Junbum Shin}, {and} \bibinfo{person}{Xiaokui Xiao}.}
  \bibinfo{year}{2018}\natexlab{}.
\newblock \showarticletitle{Privacy Enhanced Matrix Factorization for
  Recommendation with Local Differential Privacy}.
\newblock  \bibinfo{volume}{30}, \bibinfo{number}{9} (\bibinfo{year}{2018}),
  \bibinfo{pages}{1770--1782}.
\newblock


\bibitem[\protect\citeauthoryear{Wang, Zheng, Jiang, and Ren}{Wang
  et~al\mbox{.}}{2018b}]%
        {Wang201821}
\bibfield{author}{\bibinfo{person}{Cong Wang}, \bibinfo{person}{Yifeng Zheng},
  \bibinfo{person}{Jinghua Jiang}, {and} \bibinfo{person}{Kui Ren}.}
  \bibinfo{year}{2018}\natexlab{b}.
\newblock \showarticletitle{Toward Privacy-Preserving Personalized
  Recommendation Services}.
\newblock \bibinfo{journal}{\emph{Engineering}} \bibinfo{volume}{4},
  \bibinfo{number}{1} (\bibinfo{year}{2018}), \bibinfo{pages}{21--28}.
\newblock
\showISSN{2095-8099}


\bibitem[\protect\citeauthoryear{Wang, Zhang, and Zhang}{Wang
  et~al\mbox{.}}{2018a}]%
        {wang2018blockchain}
\bibfield{author}{\bibinfo{person}{Shangping Wang}, \bibinfo{person}{Yinglong
  Zhang}, {and} \bibinfo{person}{Yaling Zhang}.}
  \bibinfo{year}{2018}\natexlab{a}.
\newblock \showarticletitle{A Blockchain-Based Framework for Data Sharing With
  Fine-Grained Access Control in Decentralized Storage Systems}.
\newblock \bibinfo{journal}{\emph{{IEEE} Access}}  \bibinfo{volume}{6}
  (\bibinfo{year}{2018}), \bibinfo{pages}{38437--38450}.
\newblock


\bibitem[\protect\citeauthoryear{Warner}{Warner}{1965}]%
        {warner1965randomized}
\bibfield{author}{\bibinfo{person}{Stanley~L Warner}.}
  \bibinfo{year}{1965}\natexlab{}.
\newblock \showarticletitle{{Randomized response: A survey technique for
  eliminating evasive answer bias}}.
\newblock \bibinfo{journal}{\emph{J. Amer. Statist. Assoc.}}
  \bibinfo{volume}{60}, \bibinfo{number}{309} (\bibinfo{year}{1965}),
  \bibinfo{pages}{63--69}.
\newblock


\bibitem[\protect\citeauthoryear{Xia, Liu, Sun, and Chen}{Xia
  et~al\mbox{.}}{2015}]%
        {XiaLSC15}
\bibfield{author}{\bibinfo{person}{Peng Xia}, \bibinfo{person}{Benyuan Liu},
  \bibinfo{person}{Yizhou Sun}, {and} \bibinfo{person}{Cindy~X. Chen}.}
  \bibinfo{year}{2015}\natexlab{}.
\newblock \showarticletitle{Reciprocal Recommendation System for Online
  Dating}. In \bibinfo{booktitle}{\emph{ASONAM~'15: Proceedings of the 2015
  {IEEE/ACM} International Conference on Advances in Social Networks Analysis
  and Mining}}. \bibinfo{publisher}{{ACM}}, \bibinfo{pages}{234--241}.
\newblock


\end{thebibliography}

\end{document}